\documentclass[aps,prx,floatfix,twocolumn,letterpaper,superscriptaddress,
longbibliography,nofootinbib,
  ]{revtex4}
\usepackage[latin9]{inputenc}
\usepackage{color}
\usepackage{amssymb}
\usepackage{graphicx}
\usepackage{tikz}
\usepackage{verbatim}
\usepackage{amssymb}
\usepackage{graphicx}
\usepackage{tikz}
\usepackage{amsmath,bm,amssymb,latexsym,galois,amsthm,euscript,dsfont}
\usepackage[all,cmtip,knot]{xy}
\usepackage{dsfont} 
\usepackage{mathrsfs}
\usepackage[unicode=true,bookmarks=true,bookmarksnumbered=false,bookmarksopen=false,breaklinks=false,pdfborder={0 0 1},backref=false,colorlinks=true]{hyperref}

\hypersetup{
    colorlinks,
    linkcolor={red!50!black},
    citecolor={blue!90!black},
    urlcolor={blue!90!black}
}

\makeatletter
\@ifundefined{textcolor}{}
{%
 \definecolor{BLACK}{gray}{0}
 \definecolor{WHITE}{gray}{1}
 \definecolor{RED}{rgb}{1,0,0}
 \definecolor{GREEN}{rgb}{0,1,0}
 \definecolor{BLUE}{rgb}{0,0,1}
 \definecolor{CYAN}{cmyk}{1,0,0,0}
 \definecolor{MAGENTA}{cmyk}{0,1,0,0}
 \definecolor{YELLOW}{cmyk}{0,0,1,0}
}

\usepackage{amsfonts}\usepackage{tabularx}\usepackage{dcolumn}\usepackage{bm}\usepackage{graphicx}\usepackage{epstopdf}

\hypersetup{urlcolor=blue}

\usepackage{relsize}
\newcommand{\del}[1]{{\iffalse #1 \fi}}

\newcommand{\f}[1]{\vcenter{\hbox{\includegraphics{#1}}}}

\renewcommand{\v}[1]{\boldsymbol{#1}}

\theoremstyle{definition}
\newtheorem{defn}{Definition}

\newtheorem*{defn*}{Definition}
\newtheorem{thm}[defn]{Theorem}
\theoremstyle{remark}

\makeatother

\input{Qcircuit.tex}

\begin{document}

\title{Quantum Neural Network States: A Brief Review of Methods and Applications}
\author{Zhih-Ahn Jia}
\email{zajia@math.ucsb.edu, giannjia@foxmail.com }
\affiliation{Key Laboratory of Quantum Information, Chinese Academy of Sciences, School of Physics, University of Science and Technology of China, Hefei, Anhui, 230026, P.R. China}
\affiliation{CAS Center For Excellence in Quantum Information and Quantum Physics, University of Science and Technology of China, Hefei, Anhui, 230026, P.R. China}
\affiliation{Microsoft Station Q and Department of Mathematics, University of California, Santa Barbara, California 93106-6105, USA}
\author{Biao Yi}
\affiliation{Department of Mathematics, Capital Normal University, Beijing 100048, P.R. China}
\author{Rui Zhai}
\affiliation{Institute of Technical Physics, Department of Engineering Physics, Tsinghua University, Beijing 10084, People's Republic of China}
\author{Yu-Chun Wu}
\email{wuyuchun@ustc.edu.cn}
\affiliation{Key Laboratory of Quantum Information, Chinese Academy of Sciences, School of Physics, University of Science and Technology of China, Hefei, Anhui, 230026, P.R. China}
\affiliation{CAS Center For Excellence in Quantum Information and Quantum Physics, University of Science and Technology of China, Hefei, Anhui, 230026, P.R. China}
\author{Guang-Can Guo}
\affiliation{Key Laboratory of Quantum Information, Chinese Academy of Sciences, School of Physics, University of Science and Technology of China, Hefei, Anhui, 230026, P.R. China}
\affiliation{CAS Center For Excellence in Quantum Information and Quantum Physics, University of Science and Technology of China, Hefei, Anhui, 230026, P.R. China}
\author{Guo-Ping Guo}
\affiliation{Key Laboratory of Quantum Information, Chinese Academy of Sciences, School of Physics, University of Science and Technology of China, Hefei, Anhui, 230026, P.R. China}
\affiliation{CAS Center For Excellence in Quantum Information and Quantum Physics, University of Science and Technology of China, Hefei, Anhui, 230026, P.R. China}
\affiliation{Origin Quantum Computing, Hefei, 230026, P.R. China}

\begin{abstract}
One of the main challenges of quantum many-body physics is the exponential growth in the dimensionality of the Hilbert space with system size. This growth makes solving the Schr\"{o}dinger equation of the system extremely difficult. Nonetheless, many physical systems have a simplified internal structure that typically makes the parameters needed to characterize their ground states exponentially smaller. Many numerical methods then become available to capture the physics of the system. Among modern numerical techniques, neural networks, which show great power in approximating functions and extracting features of big data, are now attracting much interest. In this work, we briefly review the progress in using artificial neural networks to build quantum many-body states. We take the Boltzmann machine representation as a prototypical example to illustrate various aspects of the states of a neural network. We briefly review also the classical neural networks and illustrate how to use neural networks to represent quantum states and density operators. Some physical properties of the neural network states are discussed. For applications, we briefly review the progress in many-body calculations based on neural network states, the neural network state approach to tomography, and the classical simulation of quantum computing based on Boltzmann machine states.
\end{abstract}

\maketitle
\section{Introduction}
One of the most challenging problems in condensed matter physics is to find the eigenstate of a given Hamiltonian. The difficulty stems mainly from the power scaling of the Hilbert space dimension, which grows exponentially with the system size \cite{osborne2012hamiltonian,verstraete2015quantum}. To obtain a better understanding of quantum many-body physical systems beyond the mean-field paradigm and to study the behavior of strongly correlated electrons requires effective approaches to the problem. Although the dimension of the Hilbert space of the system grows exponentially with the number of particles in general, fortunately, physical states frequently have some internal structures, for example, obeying the entanglement area law, making it easier to solve problems than in the general case \cite{ORUS2014,landau2015polynomial,Arad2017,Schuch2007,Anshu2016}. Physical properties of the system usually restrict the form of the ground state, for example, area-law states \cite{Eisert2010}, ground states of local gapped systems \cite{Amico2008}. Therefore, many-body localized systems can be efficiently represented by a tensor network \cite{Friesdorf2015,ORUS2014,Verstraete2009,orus2018tensor}, which is a new tool developed in recent years to attack difficulties in representing quantum many-body states efficiently. Tensor network approach achieves some great success in quantum many-body problems. It has become a standard tool and many classical algorithm-based tensor networks have been developed, such as the density-matrix renormalization group \cite{White1992}, projected entangled pair states (PEPS) \cite{verstraete2004renormalization}, folding algorithm \cite{Banuls2009}, entanglement renormalization \cite{Vidal2007}, and time-evolving block decimation \cite{Vidal2003}. The research on tensor-network states also includes studies on finding new representations of quantum many-body states.

During the last few years, machine learning has grown rapidly as an interdisciplinary field. Machine learning techniques have also been successfully applied in many different scientific areas \cite{lecun2015deep,Hinton2006,sutton1998reinforcement}: computer vision, speech recognition, and chemical synthesis, Combining quantum physics and machine learning has generated a new exciting field of research, quantum machine learning \cite{biamonte2017quantum}, which has recently attracted much attention \cite{Rebentrost2014,Dunjko2016,Monras2017,carrasquilla2017machine,Carleo602,Deng2017,Deng2017a,gao2017efficient}. The research on quantum machine learning can be loosely categorized into two branches: developing new quantum algorithms, which share some features of machine learning and behave faster and better than their classical counterparts \cite{Rebentrost2014,Dunjko2016,Monras2017}, using classical machine learning methods to assist the study of quantum systems, such as distinguishing phases \cite{carrasquilla2017machine}, quantum control \cite{August2017}, error-correcting of topological codes \cite{Torlai2017}, and quantum tomography \cite{Zhang2017,torlai2017many}. The latter is the focus of this work. Given the substantial progress so far, we stress here that machine learning can also be used to attack the difficulties encountered with quantum many-body states.

Since 2001, researchers have been trying to use machine learning techniques, especially neural networks, to deal with the quantum problems, for example, solving the Schr\"{o}dinger equations \cite{monterola2001solving,monterola2003solving,caetano2011using,manzhos2009improved}. Later, in 2016, neural networks were introduced as a variational ansatz for representing quantum many-body ground states \cite{Carleo602}. This stimulated an explosion of results to apply machine learning methods in the investigations of condensed matter physics; see, e.g., Refs.~\cite{Rebentrost2014,Dunjko2016,Monras2017,carrasquilla2017machine,Deng2017,Deng2017a,jia2018efficient,gao2017efficient}. Carleo and Troyer initially introduced the restricted BM (RBM) to solve the transverse-field Ising model and antiferromagnetic Heisenberg model and study the time evolution of these systems \cite{Carleo602}. Later, the entanglement properties of the RBM states were investigated \cite{Deng2017}, as was their representational power \cite{gao2017efficient,huang2017neural}. Many explicit RBM constructs for different systems were given, including the Ising model \cite{Carleo602}, toric code \cite{Deng2017a}, graph states \cite{gao2017efficient}, stabilizer code \cite{jia2018efficient,zhang2018efficient}, and topologically ordered states \cite{huang2017neural,Deng2017a,jia2018efficient,lu2018efficient}. Furthermore, the deep BM (DBM) states were also investigated under different approaches \cite{gao2017efficient,Gan2017,jia2018DBM}.

Despite all the progress in applying neural networks in quantum physics, many important topics still remain to be explored. The obvious topics are the exact definition of a quantum neural network state and the mathematics and physics behind the efficiency of quantum neural network states. Although RBM and DBM states are being investigated from different aspects, there are many other neural networks. It is natural to ask if they can similarly be used for representing quantum states and what are the relationships and differences between these representations. Digging deeper, one central problem in studying neural network is its representational power. We can ask a) what is its counterpart in quantum mechanics and how to make the neural network work efficiently in representing quantum states, and b) what kind of states can be efficiently represented by a specific neural network. In this work, we investigate partially these problems and review the important progress in the field.

The work is organized as follows. In Section~\ref{sec:ANN}, we introduce the definition of artificial neural network. We explain in Section~\ref{sec:ANN1} the feed-forward neural network, perceptron and logistic neural networks, convolutional neural network, and stochastic recurrent neural network, the so-called Boltzmann machine (BM). Next, in Section~\ref{sec:ANN2} we explain the representational power for the feed-forward neural network in representing given functions and the BM in approximating given probability distributions. In Section~\ref{sec:ANNstate}, we explain how a neural network can be used as a variational ansatz for quantum states; the method given in Section~\ref{sec:ANNstate1} is model-independent, that is, the way to construct states can be applied to any neural network (with the ability to continuously output real or complex numbers). Some concrete examples of neural network states are given in Section~\ref{sec:ANNstate2}. Section~\ref{sec:ANNstate3} is devoted to the efficient representational power of neural network in representing quantum states, and in Section~\ref{sec:tensor} we introduce the basic concepts of a tensor-network state, which is closely relevant to neural network states. In Section~\ref{sec:density}, we briefly review the neural network representation of the density operator and the quantum state tomography scheme based on neural network states. In Section~\ref{sec:entanglement}, we discuss the entanglement features of the neural network states. The application of these states in classically simulating quantum computing circuit is discussed in Section~\ref{sec:QC}. In the last section, some concluding remarks are given.

\section{Artificial neural networks and their representational power}
\label{sec:ANN}
A neural network is a mathematical model that is an abstraction of the biological nerve system, which consists of adaptive units called neurons that are connected via a an extensive network of synapses \cite{Kohonen1988}. The basic elements comprising the neural network are artificial neurons, which are the mathematical abstractions of the biological neurons. When activated each neuron releases neurotransmitters to connected neurons and changes the electric potentials of these neurons. There is a threshold potential value for each neuron; while the electric potential exceeds the threshold, the neuron is activated.

\begin{figure}
  \centering
  \includegraphics[width=8cm]{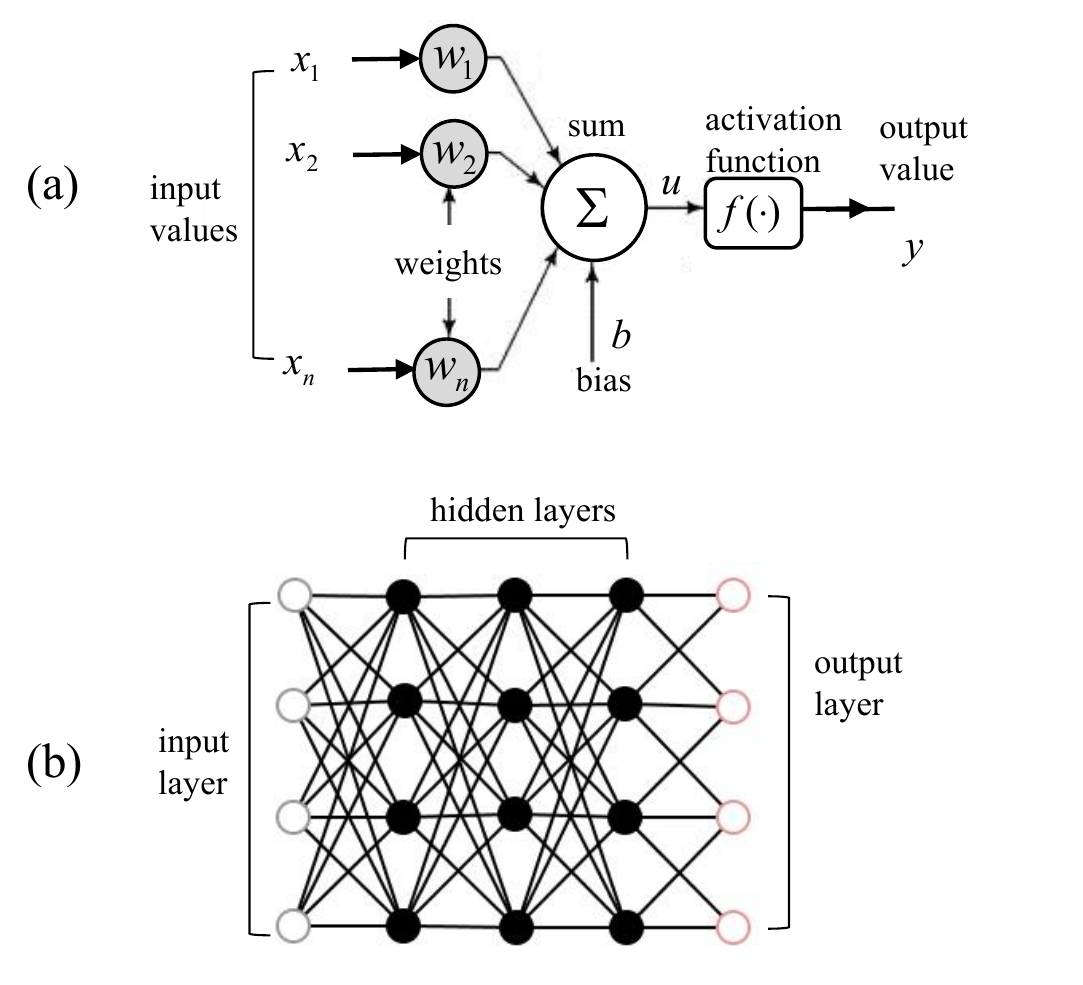}\\
  \caption{(a) McCulloch\textendash Pitts neuron model, where $x_1,\cdots,x_n$ are inputs to the neurons, $w_1,\cdots,w_n$ are weights corresponding to each input, $\Sigma$ is the summation function, $b$ the bias, $f$ the activation function, and $y$ the output of the neuron; (b) a simple artificial neural network.}\label{fig:neuron}
\end{figure}

There are several kinds of artificial neuron models. Here, we introduce the most commonly used McCulloch\textendash Pitts neuron model \cite{McCulloch1943}. Consider $n$ inputs $x_1,x_2,\cdots,x_n$, which are transmitted by $n$ corresponding weighted connections $w_1,w_2,\cdots,w_n$ (see Figure~\ref{fig:neuron}). After the signals have reached the neuron, they are added together according to their weights, and then the value is compared with the bias $b$ of the neuron to determine whether the neuron is activated or deactivated. The process is governed by the activation function $f$, and the output of a neuron is written $y=f(\sum_{i=1}^n w_ix_i-b)$. Note that we can regard the bias as a weight $w_0$ for some fixed input $x_0=-1$, the output then has a more compact form, $y=f(\sum_{i=0}^n w_ix_i)$. Putting a large number of neurons together and allowing them to connect with each other in some kind of connecting pattern produces a neural network. Note that this is just a special representation to be able to picture the neural network intuitively, especially in regard to feed-forward neural networks. Indeed, there are many other forms of mathematical structures by which to characterize the different kinds of neural networks. Here we give several important examples of neural networks.

\subsection{Some examples of neural networks}
\label{sec:ANN1}
An artificial neural network is a set of neurons where some, or all, the neurons are connected according to a certain pattern. Note that we put neurons at both input and output ends. These input and output neurons are not neurons as introduced previously, but depend on the learning problems. There may or may not be activation functions associated with them. In what follows, we briefly introduce the feed-forward, convolutional neural networks, and the Boltzmann machines.

\subsubsection{Rosenblatt's perceptron and logistic neural network}
To explain the neural network, we first see an example of the feed-forward neural network, perceptron, which was invented by Rosenblatt. In the history of artificial neural networks, the multilayer perceptron plays a crucial role. In a perceptron, the activation function of each neuron is set to a Heaviside step function
\begin{equation}\label{}
f(x)=\left\{
\begin{array}{ll}
0, &x\leq 0,\\
1, &x>0,
\end{array}\right.
\end{equation}
where one value represents the activation status of the neuron and zero value represents the deactivation status of the neuron. The output value of one neuron is the input value of the neurons connecting it.

The power of the perceptron mainly comes from its hierarchical recursive structure. It has been shown that the perceptron can be used for doing universal classical computation \cite{McCulloch1943,minsky2017perceptrons,Nielsen2015}. To see this, we note that \small $\mathsf{NAND}$ \normalsize and \small $\mathsf{FANOUT}$ \normalsize operations are universal for classical computation. In the perceptron, we still assume a \small $\mathsf{FANOUT}$ \normalsize operation works \footnote{Here, we emphasize the importance of the $\mathsf{FANOUT}$ operation, which is usually omitted from the universal set of gates in the classical computation theory. However, the operation is forbidden in quantum computation by the famous no-cloning theorem.}. We only need to show that the perceptron simulates the \small $\mathsf{NAND}$ \normalsize operation. Suppose that $x_1,x_2$ are two inputs of the neuron each with weight $-2$ and the bias of the neuron set to $-3$. When the inputs are $x_1=x_2=1$, $f(x_1w_1+x_2w_2)=0$, otherwise its output is $1$, which is exactly the output of the \small $\mathsf{NAND}$ \normalsize operation.

Although the perceptron is powerful in many applications, it still has some shortcomings. The most outstanding one is that the activation function is not continuous, a small change in the weights may produce a large change in the output of the network, this makes the learning process difficult. One way to remedy this shortcoming is to smooth out the activation function, usually, by choosing the logistic function,
\begin{equation}\label{}
f(x)=\frac{1}{1+e^{-x}}.
\end{equation}
This resulting network is usually named the logistic neural network (or sigmoid neural network). In practice, logistic neural networks are used extensively. Many problems can be solved by this network. We remark that the logistic function is chosen for convenience in the updating process of learning and many other smooth step-like functions can be chosen as an activation function. In a neural network, the activation functions need not be all the same. In Table~\ref{tab:table1}, we list some popular activation functions.
\begin{table}
\caption{\label{tab:table1}Some popular activation functions.}
\begin{ruledtabular}
\begin{tabular}{l|cc}
 &function\\
\hline
logistic function & $f(x)= \frac{1}{1+e^{-x}}$\\
tanh & $\tanh(x)= \frac{e^x-e^{-x}}{e^x+e^{-x}}$\\
cos & $\cos(x)$\\
softmax\footnote{The softmax function acts on vectors $\mathbf{x}$, which are usually used in the final layer of the neural networks.} & $\sigma(\mathbf{x})_j=\frac{e^{x_j}}{\sum_{i}e^{x_i}}$ \\
rectified linear unit & $\mathrm{ReLU}(x)=\max\{0,x\}$ \\
exponential linear unit& $\mathrm{ELU}(x)=\left\{
\begin{array}{ll}
x, &x\geq 0\\
\alpha (e^x-1), &x<0
\end{array}\right.$\\
softplus& $\mathrm{SP}(x)=\ln (e^x+1)$
\end{tabular}
\end{ruledtabular}
\end{table}

Next, we see how the neural network learns with the gradient descent method. We illustrate the learning process in the supervised machine framework using two different sets of labelled data $\mathcal{S}=\{(x_i,y_i)\}_{i=1}^N$ and $\mathcal{T}=\{(z_i,t_i)\}_{i=1}^{M}$, known as training data and test data, respectively; here $y_i$ (resp. $t_i$) is the label of $x_i$ (resp. $z_i$). Our aim is to find the weights and biases of the neural network such that the network output $y(x_i)$ (which depends on network parameters $\Omega=\{w_{ij},b_i\}$) approximates $y_i$ for all training inputs $x_i$. To quantify how well the neural network approximates the given labels, we need to introduce the cost function, which measures the difference between $y(x_i)$ and $y_i$,
\begin{equation}\label{}
C(\Omega):=C(y(x_i),y_i)=\frac{1}{2N}\sum_{i=1}^{N}\| y(x_i)-y_i\|^2,
\end{equation}
where $N$ denotes the number of data in the training set, $\Omega$ the set of network parameters $w_{ij}$ and $b_j$, and the sum runs over all data in the training set. Here, we choose the quadratic norm, therefore, the cost function is called quadratic. Now our aim is to minimize the cost function as a multivariable function of the network parameters such that $C(\Omega)\approx 0$, this can be done by the well-known gradient descent method.

The intuition behind the gradient decent method is that we can regard the cost function, in some sense, as the height of a map where the place is marked by network parameters. Our aim is to go down repeatedly from some initial place (given a configuration of the neural network) until we reach the lowest point. Formally, from some given configuration of the neural network, i.e., given parameters $w_{ij}$ and $b_i$, the gradient decent algorithm needs to compute repeatedly the gradient $\nabla C=(\frac{\partial C}{\partial w_{ij}},\frac{\partial C}{\partial b_{i}})$. The updating formulae are given by
\begin{eqnarray}
&w_{ij}\to w'_{ij}=w_{ij}-\eta\frac{\partial C}{\partial w_{ij}}, \\
&b_{i}\to b'_{i}=b_{i}-\eta\frac{\partial C}{\partial b_{i}},
\end{eqnarray}
where $\eta$ is a small positive parameter known as the learning rate.

In practice, there are many difficulties in applying gradient method to train the neural network. The modified form, the stochastic gradient descent, is usually used to speed up the training process. In the stochastic gradient method, sampling over the training set is introduced, i.e., we randomly choose $N'$ samples $\mathcal{S}'=\{(X_1,Y_1),\cdots,(X_{N'},Y_{N'})\}$ such that the average gradient of cost function over $\mathcal{S}'$ equals roughly the average gradient over the whole training set $\mathcal{S}$. Then the updating formulae are accordingly modified as
\begin{eqnarray}
&w_{ij}\to w'_{ij}=w_{ij}-\frac{\eta}{N'}\sum_{i=1}^{N'}
\frac{\partial C(X_i)}{\partial w_{ij}}, \\
&b_{i}\to b'_{i}=b_{i}-\frac{\eta}{N'}\sum_{i=1}^{N'}
\frac{\partial C(X_i)}{\partial b_{i}},
\end{eqnarray}
where $C(X_i)=\|y(X_i)-Y_i\|^2/2$ is the cost function over the training input $X_i$.

The test data $\mathcal{T}$ is usually chosen differently from $\mathcal{S}$, and when the training process is done, the test data is used to test the performance of the neural network, which for many traditionally difficult problems (such as classification and recognition) is very good. As discussed later, the feed-forward neural network and many other neural networks also work well in approximating quantum states \cite{Cai2018,saito2017solving}, this being the main theme of this paper.

\subsubsection{Convolutional neural network}
Convolutional neural networks are another important class of neural network and are most commonly used to analyze images. A typical convolutional neural network consists of a sequence of different interleaved layers, including a convolutional layer, a pooling layer, and a fully connected layer. Through a differentiable function, every layer transforms the former layer's data (usually pixels) into a new set of data (pixels).

For regular neural networks, each neuron is fully connected with the neurons in the previous layer. However, for the convolution layer of a convolutional neural network, the neurons only connect with neurons in a local neighborhood of the previous layer. More precisely, in a convolutional layer, the new pixel values of the $k$-th layer are obtained from the $(k-1)$-th layer by a filter that determines the size of the neighborhood and then gives $v^{(k)}_{ij}=\sum_{p,q}w_{ij;pq}v^{(k-1)}_{p,q}$ where the sum runs over the neurons in the local neighborhood of $v^{(k)}_{ij}$. After the filter scans the whole image (all pixel values), a new image (new set of pixel values) is obtained. The pooling layers are usually periodically added in-between successive convolutional layers and its function is to reduce the data set. For example, the max (or average) pooling chooses the maximum (or average) value of the pixels of the previous layer contained in the filter. The last fully connected layer is the same as the one in the regular neural network and outputs a class label used to determine which class the image is categorized in.

The weights and biases of the convolutional neural networks are learnable parameters, but the variables such as the size of the filter and the number of interleaved convolutional and pooling layers are usually fixed. The convolutional neural network performs well in classification-type machine learning tasks such as image recognition \cite{lecun1995convolutional,krizhevsky2012imagenet,lecun2015deep}. As has been shown numerically \cite{Liang2018}, the convolutional neural network can also be used to build quantum many-body states.

\subsubsection{Boltzmann machine}
Now we introduce another special type of artificial neural networks, the Boltzmann machine (also known as the stochastic Hopfield network with hidden units), which is an energy-based neural network model \cite{hinton1983optimal,ackley1985learning}. Recently introduced in many different physical areas \cite{Carleo602,Torlai2016,aoki2016restricted,weinstein2017learning, Huang2017,gao2017efficient,Deng2017,huang2017neural,Torlai2017,jia2018efficient,Chen2018,You2018}, the quantum versions of BMS, quantum BMs, have also been investigated \cite{Amin2018}. As the BM is very similar to the classical Ising model, here we explain the BM neural network by frequently referring to the terminology of the Ising model. Notice that the BM is very different from the perceptrons and logistic neural network as it does not treat each neuron individually. Therefore, there is no activation function attached to each specific neuron. Instead, the BM treats neurons as a whole.

Given a graph $G$ with vertex set $V(G)$ and edge set $E(G)$, the neurons $s_1,\cdots,s_n$ (spins in the Ising model) are put on vertices, $n=|V(G)|$. If two vertices $i$ and $j$ are connected, there is a weight $w_{ij}$ (coupling constant in the Ising model) between the corresponding neurons $s_i$ and $s_j$. For each neuron $s_i$, there is also a corresponding local bias (local field in the Ising model). As has been done for Ising model, for each series of input values $\mathbf{s}=(s_1,\cdots,s_n)$ (spin configuration in the Ising model), we define its energy as
\begin{equation}\label{}
E(\mathbf{s})=-\sum_{\langle ij\rangle \in E(G)} w_{ij}s_is_j-\sum_i s_ib_i.
\end{equation}
Up to now, everything is just as for the Ising model. No new concepts or techniques are introduced. The main difference is that, the BM construction introduces a coloring on each vertex. Each vertex receives a label \emph{hidden} or \emph{visible}. We assume the first $k$ neurons are hidden neurons denoted by $h_1,\cdots,h_k$, and the left $l$ neurons are visible neurons denoted by $v_1,\cdots,v_l$ and $k+l=n$. Therefore, the energy is now $E(\mathbf{h},\mathbf{v})$. The BM is a parametric model of a joint probability distribution between variables $\mathbf{h}$ and $\mathbf{v}$ with the probability given by
\begin{equation}\label{}
p(\mathbf{h},\mathbf{v})=\frac{e^{-E(\mathbf{h},\mathbf{v})}}{Z},
\end{equation}
where $Z=\sum_{\mathbf{h},\mathbf{v}}e^{-E(\mathbf{h},\mathbf{v})}$ is the partition function.

The general BM is very difficult to train, and therefore some restricted architecture on the BM is introduced. The restricted BM (RBM) was initially invented by Smolensky~\cite{smolensky1986information} in 1986. In the RBM, it is assumed that the graph $G$ is a bipartite graph; the hidden neurons only connect with visible neurons and there are no intra-layer connections. This kind of restricted structure makes the neural network easier to train and therefore has been extensively investigated and used \cite{Carleo602,Torlai2016,aoki2016restricted,weinstein2017learning,Huang2017,gao2017efficient,Deng2017,huang2017neural,Torlai2017,jia2018efficient,Chen2018,You2018}. The RBM can approximate every discrete probability distribution \cite{Le2008,Montufar2011}.

The BM is most notably a stochastic recurrent neural network whereas the perceptron and the logistic neural network are feed-forward neural networks. There are many other types of neural networks. For a more comprehensive list, see textbooks such as Refs.~\cite{bishop1995neural,fausett1994fundamentals}. The BM is crucial in quantum neural network states and hence its neural network states are also the most studied. In later sections, we shall discuss the physical properties of the BM neural network states and their applications.

\subsubsection{Tensor networks}
Tensor networks are certain contraction pattern of tensors, which play an important role in many scientific areas such as condensed matter physics, quantum information and quantum computation, computational physics, and quantum chemistry \cite{ORUS2014,landau2015polynomial, Arad2017,Schuch2007,Anshu2016,orus2018tensor}. We discuss some details of tensor networks latter in this review. Here, we only comment on the connection between tensor networks and machine learning.

Many different tensor network structures have been developed over the years for solving different problems like the matrix product states (MPS) \cite{fannes1992finitely,klumper1993,klumper1991equivalence}, projective entangled pair states (PEPS) \cite{verstraete2004renormalization}, multiscale entanglement renormalization ansatz (MERA) \cite{Vidal2007}, branching \cite{Evenbly2014,Evenbly2014Scaling}, and tree tensor network \cite{Shi2006}, matrix product operator \cite{Zwolak2004,Cui2015,Gangat2017,Chen2018exponetial}, projective entangled pair operator \cite{Czarnik2015,Parish2016,kshetrimayum2017simple}, and continuous tensor networks \cite{Verstraete2010,Haegeman2011,Haegeman2013}. A large number of numerical algorithms based on tensor networks are now available, including the density-matrix renormalization group \cite{White1992}, folding algorithm \cite{Banuls2009}, entanglement renormalization \cite{Vidal2007}, time-evolving block decimation \cite{Vidal2003}, and tangent space method \cite{Haegeman2013a}.

One of the most important properties that empowers tensor networks is that entanglement is much easier to treat in this representation. Many studies have appeared in recent years that indicate that tensor networks have a close relationship with state-of-the-art neural network architectures. From theory,  machine learning architectures were shown in Ref.~\cite{levine2018bridging} to be understood via the tensor networks and their entanglement pattern. In practical applications, tensor networks can also be used for many machine-learning tasks, for example, performing learning tasks by optimizing the MPS \cite{stoudenmire2016supervised,Han2018}, preprocessing the dataset based on layered tree tensor networks \cite{stoudenmire2018learning}, classifying images via the MPS and tree tensor networks \cite{stoudenmire2016supervised,stoudenmire2018learning, Han2018,liu2017machine}, and realizing quantum machine learning via tensor networks \cite{huggins2018towards}. Both tensor networks and neural networks can be applied to represent quantum many-body states; the difference and connections of the two kinds of representations are extensively explored in several works \cite{Chen2018,Glasser2018,Deng2017, gao2017efficient,jia2018DBM,jia2018efficient,zhang2018efficient},. We shall review some of these progress in detail in Section~\ref{sec:ANNstate}.

\subsection{Representational power of neural network}
\label{sec:ANN2}
Next we comment on the representational power of neural networks, which is important in understanding the representational power of quantum neural network states. In 1900, Hilbert formulated his famous list of 23 problems, among which the thirteenth problem is devoted to the possibility of representing an $n$-variable function as a superposition of functions of a lesser number of variables. This problem is closely related to the representational power of neural networks. Kolmogorov \cite{kolmogorov1956representation,kolmogorov1957representation} and Arnold \cite{arnold2009vladimir} proved that for continuous $n$-variable functions, this is indeed the case. The result is known as the Kolmogorov\textendash Arnold representation theorem (alternatively, the Kolmogorov superposition theorem);
\begin{thm}
Any $n$-variable real continuous function $f:[0,1]^n\to \mathbb{R}$ expands as sums and compositions of continuous univariate functions; more precisely, there exist real positive numbers $a,b,\lambda_p, \lambda_{p,q}$ and a real monotonic increasing function $\phi:[0,1]\to [0,1]$ such that
\begin{equation}\label{eq:KA}
f(x_1,\cdots,x_n)=\sum_{q=1}^{2n+1}F(\sum_{p=1}^n  \lambda_p \phi(x_p+ aq)+bq),
\end{equation}
or
\begin{equation}\label{eq:KA}
f(x_1,\cdots,x_n)=\sum_{q=1}^{2n+1}F(\sum_{p=1}^n  \lambda_{pq} \phi(x_p+ aq)),
\end{equation}
where $F$ is a real and continuous function called the outer function, and $\phi$ called the inner function. Note that $a$ and $F$ may be different in two representations.
\end{thm}

Obviously, the mathematical structure in the theorem is very similar to the mathematical structure of feed-forward neural networks. Since the initial work of Kolmogorov and Arnold, numerous follow-up work contributed to understanding more deeply the representation power of neural networks from different aspects \cite{alexeev2010neural}. Mathematicians have considered the problem in different support sets and different metric between functions. The discrete-function version of the problem is also extensively investigated. As mentioned in the previous section, McCulloch and Pitts \cite{McCulloch1943} showed that any Boolean function can be represented by the perceptron and, based on this fact, Rosenblatt developed the learning algorithm \cite{rosenblatt1961principles}. S{\l}upecki proved that all $k$-logic functions can be represented as a superposition of one-variable functions and any given significant function \cite{slupecki1972criterion}. Cybenko \cite{cybenko1989approximations}, Funahashi \cite{funahashi1989approximate} and Hornik and colleagues \cite{hornik1989multilayer} proved that $n$-variable functions defined on a compact subset of $\mathbb{R}^n$ may be approximated by a four-layer network with only logistic activation functions and a linear activation function. Hecht \cite{hecht1987kolmogorov} went a step further; he proved that any $n$-variable continuous function can be represented by a two-layer neural network involving logistic activation functions of the first layer and arbitrary activation functions on the second layer. These results are summarized as follows;
\begin{thm}
The feed-forward neural network can approximate any continuous $n$-variable functions and any $n$-variable discrete functions.
\end{thm}

For the stochastic recurrent neural network BM, the power in approximating probability distributions has also been studied extensively. An important result of Le Roux and Bengio \cite{Le2008} claims that
\begin{thm}
Any discrete probability distribution $p:\mathbb{B}^n\to \mathbb{R}_{\geq 0}$ can be approximated with an RBM with $k + 1$ hidden neurons where $k=|\mathrm{supp}(p)|$ is the cardinality of the support of $p$ (i.e., the number of vectors with non-zero probabilities) arbitrarily well in the metric of the Kullback\textendash Leibler divergence.
\end{thm}
The theorem states that any discrete probability distribution can be approximated by the RBM. The bound of the number of hidden neurons is later improved \cite{Montufar2011}.

Here we must stress that these representation theorems are applicable only if the given function or probability distribution can be represented by the neural network. In practice, the number of parameters to be learned cannot be too large for the number of input neurons when we build a neural network. If a neural network can represent a function or distribution in polynomial time (the number of parameters depends polynomially on the number of input neurons), we say that the representation is efficient.

\section{Artificial neural network ansatz for quantum many-body system}
\label{sec:ANNstate}

\subsection{Neural network ansatz state}
\label{sec:ANNstate1}
\begin{figure}
  \centering
  \includegraphics[width=4cm]{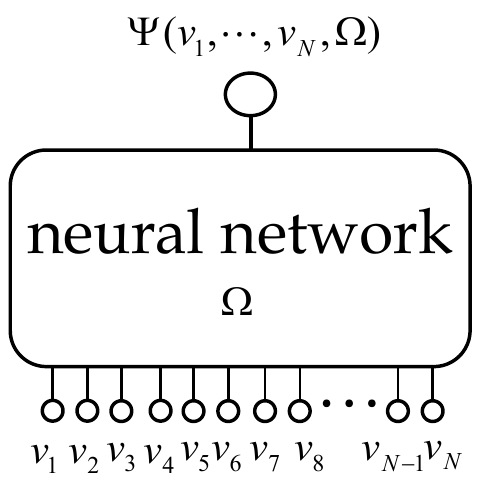}\\
  \caption{Schematic diagram for the neural network ansatz state.}\label{fig:state}
\end{figure}

We now describe how neural network can be used as a variational ansatz for quantum many-body systems. For a given many-body pure state $|\Psi\rangle$ of an $N$-particle $p$-level physical system
$$|\Psi\rangle=\sum_{v_1,v_2\cdots,v_N=1}^p\Psi(v_1,v_2\cdots,v_N)
|v_1\rangle\otimes|v_2\rangle\otimes\cdots\otimes|v_N\rangle.$$
The coefficient $\Psi(v_1,v_2\cdots,v_N)$ of the state can be regarded as an $N$-variable complex function. To characterize a state, we only need to give the corresponding value of $\Psi$ function for each variable $\mathbf{v}=(v_1,v_2\cdots,v_N)$. One of the difficulties in quantum many-body physics is that the complete characterization of an $N$-particle system requires $O(p^N)$ coefficients, which is exponentially large in the system size $N$ and therefore is computationally inefficient. Let us now see how neural network can be used to attack the difficulty.

To represent a quantum state, we first need to build a specific architecture of the neural network for which we denote the set of adjustable parameters as $\Omega=\{w_{ij},b_i\}$. The number of input neurons is assumed to be the same as the number of physical particles $N$. For each series of inputs $\mathbf{v}=(v_1,\cdots,v_N)$, we anticipate the neural network to output a complex number $\Psi(\mathbf{v},\Omega)$, which depends on values of both the input and parameters of the neural network. In this way, a variational state
\begin{equation}\label{eq:NNS}
|\Psi(\Omega)\rangle = \sum_{\mathbf{v}\in \mathbb{Z}_p^N}
\Psi(\mathbf{v},\Omega)|\mathbf{v}\rangle,
\end{equation}
is obtained, where the sum runs over all basis labels, $|\mathbf{v}\rangle$ denotes $|v_1\rangle\otimes\cdots\otimes|v_N\rangle$, and $\mathbb{Z}_p$ the set $\{0,1,\cdots,p-1\}$ (the labels of local basis). The state in Equation~(\ref{eq:NNS}) is a variational state. For a given Hamiltonian $H$, the corresponding energy functional is
\begin{equation}\label{eq:energyfunctional}
E(\Omega)=\frac{\langle\Psi(\Omega)|H|\Psi(\Omega)\rangle}
{\langle\Psi(\Omega)|\Psi(\Omega)\rangle}.
\end{equation}
In accordance with the variational method, the aim now is to minimize the energy functional and obtain the corresponding parameter values, with which the (approximate) ground state is obtained. The process of adjusting parameters and finding the minimum of the energy functional is performed using neural network learning (see Figure~\ref{fig:state}). Alternatively, if the appropriate dataset exists, we can also build the quantum neural network states by standard machine learning procedures rather than minimizing the energy functional. We first build a neural network with learnable parameters $\Omega$ and then train the network with the available dataset. Once the training process is completed, the parameters of the neural network are fixed; we also obtain the corresponding approximate quantum states.

The notion of the efficiency of the neural network ansatz in representing a quantum many-body state is defined as the dependency relation of the number of non-zero parameters $|\Omega|$ involved in the representation and the number of physical particles $N$: if $|\Omega|=O(\mathsf{poly}(N))$, the representation is called efficient. The aim when solving a given eigenvalue equation is therefore to build a neural network for which the ground state can be represented efficiently.

To obtain the quantum neural network states from the above construction, we first need to make the neural network a complex neural network, specifically, use complex parameters and output complex values. In practice, some neural networks may have difficulty outputing complex values. Therefore, we need to develop another way to build a quantum neural network state $|\Psi\rangle$. We know that wavefunction $\Psi(\mathbf{v})$ can be written as $\Psi(\mathbf{v})=R(\mathbf{v})e^{i\theta(\mathbf{v})}$ where the amplitude $R(\mathbf{v})$ and phase $\theta(\mathbf{v})$ are both real functions; hence, we can represent them by two separate neural networks with parameter sets $\Omega_1$ and $\Omega_2$. The quantum states are determined from the union of these sets, $\Omega=\Omega_1\cup \Omega_2$ (Figure~\ref{fig:density}). In Section~\ref{sec:density}, we give an explicit example to clarify the construction. Hereinafter, most of our discussion remains focused on the complex neural network approach.

Let us now see some concrete examples of neural network states.
\subsubsection{Some examples of neural networks states}
\label{sec:ANNstate2}
\begin{figure}
  \centering
  \includegraphics[width=8cm]{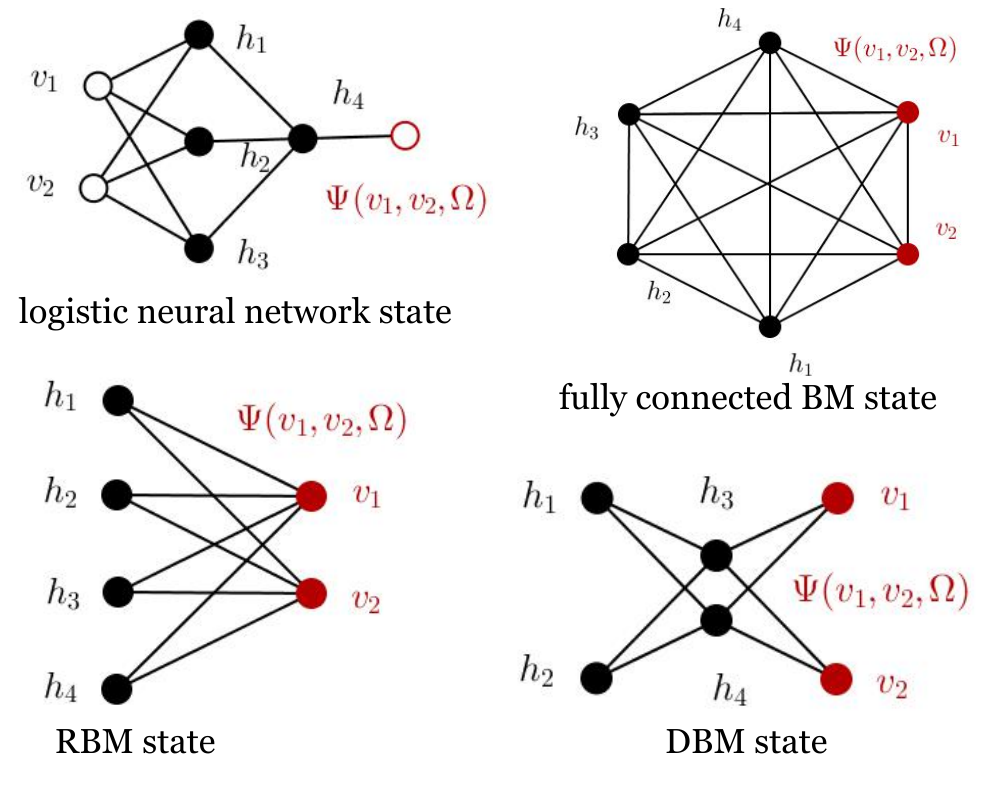}\\
  \caption{Examples of two-qubit neural network ansatz states.}\label{fig:example}
\end{figure}

The first neural network state we consider is the logistic neural network state, where weights and biases now must be chosen as complex numbers and the activation function $f(z)=1/(1+e^{-z})$ is also a complex function. As shown in Figure~\ref{fig:example}, we take the two-qubit state as an example. We assume the biases are $b_1,\cdots,b_4$ for hidden neurons $h_1,\cdots,h_4$ respectively; the weights between neurons are denoted by $w_{ij}$. We construct the state coefficient neuron by neuron next.

In Figure~\ref{fig:example}, the output for $h_i$, $i=1,2,3$ is $y_i=f(v_1w_{1i}+v_2w_{2i} -b_i)$, respectively. These outputs are transmitted to $h_4$; after acting with $h_4$, we get the state coefficient,
\begin{equation}\label{}
\Psi_{\mathrm{log}}(v_1,v_2,\Omega)=f(w_{14}y_1+w_{24}y_2+w_{34}y_3-b_4),
\end{equation}
where $\Omega=\{w_{ij},b_i\}$. Summing over all possible input values, we obtain the quantum state $|\Psi_{\mathrm{log}}(\Omega)\rangle=\sum_{v_1,v_2}\Psi_{\mathrm{log}}(v_1,v_2, \Omega)|v_1,v_2\rangle$ up to a normalization factor. We see that the logistic neural network states have a hierarchical iteration control structure that is responsible for the representation power of the network in representing states.

However, when we want to give the neural network parameters of a given state $|\Psi\rangle$ explicitly, we find that $f(z)=1/(1+e^{-z})$ cannot exactly take values zero and one as they are the asymptotic values of $f$. This shortcoming can be remedied by a smoothing step function in another way. Here we give a real function solution; the complex case can be done similarly. The idea is very simple. We cut the function into pieces and then glue them together in some smooth way. Suppose that we want to construct a smooth activation function $F(x)$ such that
\begin{equation}\label{eq:smooth}
F(x)\left\{
\begin{array}{lll}
=0, &x\leq -\frac{a}{2},\\
\in (0,1) & -\frac{a}{2}< x < \frac{a}{2},\\
=1, &x\geq \frac{a}{2},
\end{array}\right.
\end{equation}
we can choose a kernel function
\begin{equation}\label{}
K(x)=\left\{
\begin{array}{ll}
\frac{4x}{a^2}+\frac{2}{a}, &-\frac{a}{2}\leq x \leq 0,\\
\frac{2}{a}- \frac{4x}{a^2},& 0\leq x < \frac{a}{2},
\end{array}\right.
\end{equation}
The required function can then be constructed as
\begin{equation}\label{}
F(x)=\int_{x-\frac{a}{2}}^{x+\frac{a}{2}}K(x-t)s(t) dt,
\end{equation}
where $s(t)$ is step function. It is easy to check that the constructed function $F(x)$ is differentiable and satisfies Equation~(\ref{eq:smooth}). In this way, the explicit neural network parameters can be obtained for any given state.

Note that the above representation of the quantum state $\Psi(\mathbf{v})$ by a neural network is to develop the complex-valued neural network. It will be difficult in some cases. Because the quantum state $\Psi(\mathbf{v})$ can also be expressed as an amplitude $R(\mathbf{v})$ and phase $e^{i\theta(\mathbf{v})}$ as $\Psi(\mathbf{v})=R(\mathbf{v})e^{i\theta(\mathbf{v})}$, we can also represent the amplitude and phase by two neural networks separately as $R(\Omega_1,\mathbf{v})$ and $\theta(\Omega_2,\mathbf{v})$ where $\Omega_1$ and $\Omega_2$ are two respective parameter sets of the neural networks. The approach is used in representing a density operator by purification; to be discussed in Section~\ref{sec:density}.

For the BM states, we notice that the classical BM networks can approximate a discrete probability distribution. The quantum state coefficient $\Psi(\mathbf{v})$ is the square root of the probability distribution and therefore should also be able to be represented by the BM. This is one reason that the BM states are introduced as a representation of quantum states. Here we first treat instances of fully connected BM states (Figure~\ref{fig:example}). Instances for the RBM and DBM are similar. As in the logistic states, the weights and biases of the BM are now complex numbers. The energy function is defined as
\begin{align}\label{}
E(\mathbf{h},\mathbf{v}) =
&-(\sum_iv_ia_i+\sum_jh_jb_j+\sum_{\langle ij\rangle}w_{ij}v_ih_j \nonumber \\
& +\sum_{\langle jj'\rangle}w_{jj'}h_jh_{j'}+\sum_{\langle ii'\rangle}w_{ii'}v_iv_{i'}),
\end{align}
where $a_i$ and $b_j$ are biases of visible neurons and hidden neurons, respectively; $w_{ij}$, $w_{jj'}$, and $w_{ii'}$ are connection weights. The state coefficients are now
\begin{equation}\label{}
\Psi_{BM}(\mathbf{v},\Omega)=\sum_{h_1}\cdots\sum_{h_l}
\frac{e^{-E(\mathbf{h},\mathbf{v})}}{Z}
\end{equation}
with $Z=\sum_{\mathbf{v},\mathbf{h}}e^{-E(\mathbf{h},\mathbf{v})}$ the partition function, and the sum runs over all possible values of the hidden neurons. The quantum state is $|\Psi_{BM}(\Omega)\rangle=\frac{\sum_{\mathbf{v}}\Psi_{BM}(\mathbf{v},\Omega)|\mathbf{v}\rangle}{\mathcal{N}}$ where $\mathcal{N}$ is the normalizing factor.

Because the fully connected BM states are extremely difficult to train in practice, the more commonly used ones are the RBM states where there is one hidden layer and one visible layer. There are no intra-layer connections [hidden (resp. visible) neurons do not connect with hidden (resp. visible) neurons]. In this instance, the energy function becomes
\begin{align}\label{}
E(\mathbf{h},\mathbf{v})
  &=-\sum_{i}a_iv_i-\sum_{j}b_jh_j-\sum_{ij}v_iW_{ij}h_j.\nonumber \\
  &=-\sum_{i}a_iv_i-\sum_j h_j(b_j+\sum_iv_iW_{ij}).
\end{align}
Then the wavefunction is
\begin{align}\label{}
  \Psi(\mathbf{v},\Omega)&\sim\sum_{h_1}\cdots\sum_{h_l}
e^{\sum_{i}a_iv_i+\sum_j h_j(b_j+\sum_iv_iW_{ij})}, \nonumber \\
  &=\prod_ie^{a_iv_i}\prod_{j}\Gamma_j(\mathbf{v};b_j,W_{ij}),
\end{align}
where by `$\sim$' we mean that the overall normalization factor and the partition function $Z(\Omega)$ are omitted, $\Gamma_j=\sum_{h_j}e^{h_j(b_j+\sum_iv_iW_{ij})}$ is $2\cosh(b_j+\sum_iv_iW_{ij})$ or $1+e^{b_j+\sum_iv_iW_{ij}}$ for $h_j$ takes values in $\{\pm 1\}$ and $\{0,1\}$, respectively. This kind of product form of the wavefunction plays an important role in understanding the RBM states.

The DBM has more than one hidden layer; indeed, as has been shown in Ref.~\cite{gao2017efficient}, any BM can be transformed into a DBM with two hidden layers. Hence, we shall only be concerned with the DBM with two hidden layers. The wavefunction is written explicitly as
\begin{equation}\label{}
\Psi(\mathbf{v},\Omega)\sim\sum_{h_1}\cdots\sum_{h_l}\sum_{g_1}\cdots
\sum_{g_q}\frac{\exp^{-E(\mathbf{v},\mathbf{h},\mathbf{g})}}{Z},
\end{equation}
where the energy function is now of the form $E(\mathbf{v},\mathbf{h},\mathbf{g})=-\sum_i v_i a_i-\sum_k c_k g_k -\sum_j h_jb_j-\sum_{i,j;\langle ij\rangle}W_{ij} v_ih_j -\sum_{jk;\langle kj\rangle}W_{kj}h_j g_k$. It is also difficult to train the DBM, in general, but the DBM states have a stronger representational power than the RBM states; the details are discussed in the next subsection.

\subsubsection{Representational power of neural network states}
\label{sec:ANNstate3}

As the neural network states were introduced in many-body physics to represent the ground state of the transverse-field Ising model and the antiferromagnetic Heisenberg model efficiently \cite{Carleo602}, many researchers have studied their representation power. We now know that RBMs are capable of representing many different classes of states \cite{gao2017efficient,Deng2017a,Huang2017,jia2018efficient}. Unlike their unrestricted counterparts, RBMs allow an efficient sampling and they are also the most studied cases. The DBM states are also explored in various works \cite{gao2017efficient,Gan2017,jia2018DBM}. In this section, we briefly review the progress in this direction.

We first list some known classes of states that can be efficiently represented by RBM: $\mathbb{Z}_2$-toric code states \cite{Deng2017a}; graph states \cite{gao2017efficient}; stabilizer states with generators of pure type, $\mathbf{S}_{X},\mathbf{S}_{Y},\mathbf{S}_{Z}$ and their arbitrary union \cite{jia2018efficient}; perfect surface code states, surface code states with boundaries, defects, and twists \cite{jia2018efficient}; Kitaev's $D(\mathbb{Z}_d)$ quantum double ground states \cite{jia2018efficient}; the arbitrary stabilizer code state \cite{zhang2018efficient}; ground states of the double semion model and the twisted quantum double models \cite{lu2018efficient}; states of the Affleck--Lieb--Kennedy--Tasaki model and the two-dimensional CZX model \cite{lu2018efficient}; states of Haah's cubic code model \cite{lu2018efficient}; and the generalized-stabilizer and hypergraph states \cite{lu2018efficient}. The algorithmic way to obtain the RBM parameters of the stabilizer code state for arbitrary given stabilizer group $\mathbf{S}$ has also been developed \cite{zhang2018efficient}.

Although many important classes of states may be represented by the RMB, there is a crucial result regarding a limitation: \cite{gao2017efficient} there exist states that can be expressed as PEPS \cite{Gao2017quantum} but cannot be efficiently represented by a RBM; moreover, the class of RBM states is not closed under unitary transformations. One way to remedy the defect is by adding one more hidden layer, that is, using the DBM.

The DBM can efficiently represent physical states including:
\begin{itemize}
  \item Any state which can be efficiently represented by RBMs \footnote{This can be done by setting all the parameters involved in the deep hidden layer to zeros; only the parameters of the shallow hidden layer remain nonzero.};
  \item Any $n$-qubit quantum states generated by a quantum circuit of depth $T$; the number of hidden neurons is $O(nT)$ \cite{gao2017efficient};
  \item Tensor network states consist of $n$-local tensors with bound dimension $D$ and maximum coordination number $d$; the number of hidden neurons is $O(nD^{2d})$ \cite{gao2017efficient};
  \item The ground states of Hamiltonians with gap $\Delta$; the number of hidden neurons is $O(\frac{m^2}{\Delta}(n-\log\epsilon))$ where $\epsilon$ is the representational error \cite{gao2017efficient};
\end{itemize}
Although there are many known results concerning the BM states, the same for other neural networks nevertheless has been barely explored.

\subsection{Tensor network states}
\label{sec:tensor}
Let us now introduce a closely related representation of the quantum many-body states\textemdash the tensor network representation, which was originally developed in the context of condensed matter physics based on the idea of the renormalization group. Tensor network states have now applications in many different scientific fields. Arguably, the most important property of the tensor network states is that entanglement is much easier to read out than other representations.

Although there are many different types of tensor networks, we focus here on the two simplest and easily accessible ones, the MPS and the PEPS. For other more comprehensive reviews, see \cite{ORUS2014,landau2015polynomial,Arad2017,Schuch2007,Anshu2016,orus2018tensor}.

By definition, a rank-$n$ tensor is a complex variable with $n$ indices, for example $A_{i_1,i_2,\cdots,i_n}$. The number of values that an index $i_k$ can take is called the bond dimension of $i_k$. The contraction of two tensors is a new tensor, that being defined as the sum over any number of pairs of indices; for example, $C_{i_1,\cdots,i_p,k_1,\cdots,k_q}=\sum_{j_1,\cdots ,j_l} A_{i_1,\cdots,i_p,j_1,\cdots ,j_l} B_{j_1,\cdots ,j_l,k_1,\cdots,k_q}$. A tensor network is a set of tensors for which some (or all) of the indices are contracted.

\begin{table*}
\caption{\label{tab:table2}Some popular tensor network structures and their properties.}
\begin{ruledtabular}
\begin{tabular}{c|cccc}
Tensor network structure & Entanglement entropy $S(\mathcal{A})$
&correlation length $\xi$ &local observable $\langle \hat{O} \rangle$ & diagram\\ \hline
Matrix product state & $O(1)$ & finite & exact& $\f{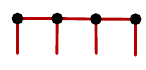}$\\
Projective entangled pair state ($2d$) & $O(|\partial \mathcal{A}|)$ & finite/infinite
& approximate & $\f{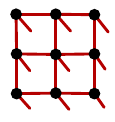}$\\
Multiscale entanglement\\ renormalization ansatz ($1d$) & $O(\log |\partial \mathcal{A}|)$
& finite/infinite & exact &$\f{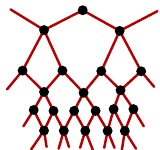}$\\
Branching multiscale entanglement \\ renormalization ansatz ($1d$)
& $O(\log |\partial \mathcal{A}|)$ & finite/infinite & exact&$\f{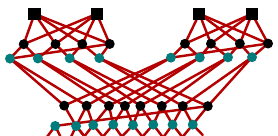}$\\
Tree tensor networks & $O(1)$ & finite & exact & $\f{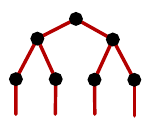}$
\end{tabular}
\end{ruledtabular}
\end{table*}

Representing the tensor network graphically is quite convenient. The corresponding diagram is called a tensor network diagram, in which, a rank-$n$ tensor is represented as a vertex with $n$-edges, for example, a scalar is just a vertex, a vector is a vertex with one edge, and a matrix is a vertex with two edges:
\begin{equation}
\mathrm{scalar:}\,\,\, \f{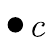};\,\,\,\mathrm{vector:}\,\,\, \f{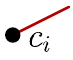};
\,\,\,\mathrm{matrix:}\,\,\, \f{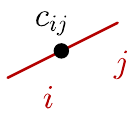}.
\label{tensors}
\end{equation}
The contraction is graphically represented by connecting two vertices with the same edge label. For two vectors and matrices, this corresponds to the inner product and the matrix product, respectively. Graphically, they look like
\begin{equation}
\mathrm{inner\ product:}\,\,\,\sum_ia_ib_i= \f{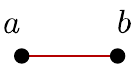};
\label{eq:inner}
\end{equation}
\begin{equation}
\mathrm{matrix\ product:}\,\,\,\sum_{j}A_{ij}B_{jk}= \f{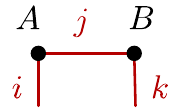}.
\label{eq:contraction}
\end{equation}

How can we use the tensor network to represent a many-body quantum state? The idea is to regard the wavefunction $\Psi(v_1,\cdots,v_n)=\langle\mathbf{v}|\Psi\rangle$ as a rank-$n$ tensor $\Psi_{v_1,\cdots,v_n}$. In some cases, the tensor wavefunction can break into some small pieces, specifically, contraction of some small tensors. For example $\Psi_{v_1,\cdots,v_n} =\sum_{\alpha_1,\cdots,\alpha_n}A^{[1]}_{i_1;\alpha_n\alpha_1}A^{[2]}_{i_2;\alpha_1\alpha_2}\cdots A^{[n]}_{i_n;\alpha_{n-1}\alpha_{n}}$. Graphically, we have
\begin{equation}
\f{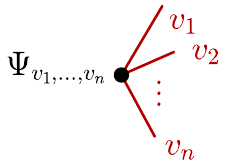}=\f{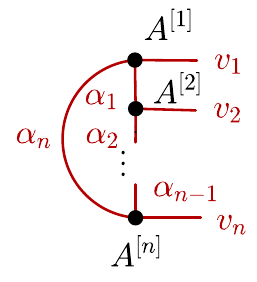},
\label{eq:wavetensor}
\end{equation}
where each $A^{[k]}_{i_k;\alpha_{k-1}\alpha_k}$ is a local tensor depending only on some subset of indices $\{v_1,\cdots,v_n\}$. In this way, physical properties such as entanglement are encoded into the contraction pattern of the tensor network diagram. It turns out that this kind of representation is very powerful in solving many physical problems.

There are several important tensor network structures. We take two prototypical tensor network states used for $1d$ and $2d$ systems, MPS states, \cite{fannes1992finitely,klumper1993,klumper1991equivalence} and PEPS states \cite{verstraete2004renormalization}, as examples to illustrate the construction of tensor-network states. In Table~\ref{tab:table2}, we list some of the most popular tensor-network structures including MPS, PEPS, MERA \cite{Vidal2007}, branching MERA \cite{Evenbly2014,Evenbly2014Scaling}, and tree tensor networks \cite{Shi2006}, We also list the main physical properties of these structures, such as correlation length and entanglement entropy. For more examples, see Refs.~\cite{ORUS2014,landau2015polynomial,Arad2017,Schuch2007, Anshu2016,orus2018tensor}

A periodic-boundary-condition MPS state is just like the right-hand side of Equation~(\ref{eq:wavetensor}), which consists of many local rank-3 tensors. For the open boundary case, the boundary local tensor is replaced with rank-2 tensors, and the inner part remains the same. The MPSs correspond to the low energy eigenstates of local gapped $1d$ Hamiltonians \cite{Hastings2006,Hastings2007}. The correlation length of the MPS is finite and they obey the entanglement area law, thus they cannot be used for representing quantum states of critical systems that break the area law \cite{Eisert2010}.

The PEPS state can be regarded as a higher-dimensional generalization of MPS. Here we give an example of a $2d$ $3\times 3$ PEPS state with open boundary
\begin{equation}
\Psi_{\mathrm{PEPS}}(\mathbf{v})=\f{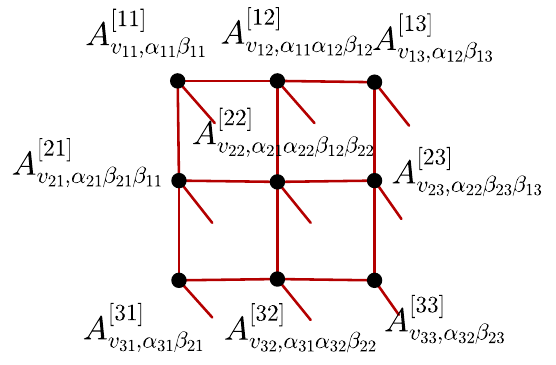}.
\label{eq:PEPS}
\end{equation}
The typical local tensors for PEPS states are rank-5 tensors for the inner part, rank-4 tensors for the boundary part and rank-3 tensors for the corner part. The $2d$ PEPSs capture the low-energy eigenstates of $2d$ local Hamiltonians, which obey the entanglement area law \cite{Eisert2010}. PEPSs have some difference with MPS; their correlation length is not always finite and can be used to represent quantum states of critical systems. However, there is, by now, no efficient way to contract physical information from PEPS exactly, therefore, many approximate methods have been developed in recent years.

The tensor network states have a close relationship with neural network states. Their connections are extensively explored in many studies \cite{Chen2018,gao2017efficient,huang2017neural}. Here, we briefly discuss how to transform a RBM state into a tensor network state. To do this, we need to regard visible and hidden neurons as tensors. For example, the visible neuron $v_i$ and hidden neuron $h_j$ is now replaced by
\begin{equation}\label{}
V^{(i)}=\left(\begin{array}{cc}1 & 0 \\0 & e^{a_i}\end{array}\right),
\end{equation}
\begin{equation}\label{}
H^{(j)}=\left(\begin{array}{cc}1 & 0 \\0 & e^{b_j}\end{array}\right),
\end{equation}
and the weighted connection between $v_i$ and $h_j$ is now also replaced by a tensor
\begin{equation}\label{}
W^{(ij)}=\left(\begin{array}{cc}1 & 1 \\1 & e^{w_{ij}}\end{array}\right).
\end{equation}
It is easy to check that both RBM and tensor network representations give the same state. Note that by some further optimization, any local RBM state can be transformed into an MPS state \cite{Chen2018}. The general correspondence between RBM and tensor-network states has been discussed in Ref.~\cite{Chen2018}. One crucial thing is that here we are only concerned with reachability, specifically, whether one representation can be represented by another. However, in practical applications, we must also know the efficiency to represent one by the other. As indicated in Section~\ref{sec:ANNstate3}, there exist some tensor network states which cannot be efficiently represented by RBM.

We note that there are also several studies trying to combine the respective advantages of a tensor network and a neural network to give a more powerful representation of the quantum many-body states \cite{Glasser2018}.

\subsection{Advances in quantum many-body calculations}
There are several studies concerning numerical tests of the accuracy and efficiency of neural network states for different physical systems and different physical phenomena \cite{monterola2001solving,monterola2003solving,caetano2011using,manzhos2009improved,Carleo602,Rebentrost2014,Dunjko2016,Monras2017,carrasquilla2017machine,Deng2017,Deng2017a,jia2018efficient,gao2017efficient}. The early work trying to use a neural network to solve the Schr\"{o}dinger equations \cite{monterola2001solving,monterola2003solving, caetano2011using,manzhos2009improved} date back to 2001. Recently, in 2016, Carleo and Troyer made the approach popular in calculating physical quantities of the quantum systems \cite{Carleo602}. Here we briefly discuss several examples of numerical calculations in many-body physics, including spin systems, and bosonic and fermionic systems.

\paragraph{Transverse-field Ising model.}\textemdash The Hamiltonian for the Ising model immersed in a transverse field is given by
\begin{equation}\label{}
H_{tIsing}=-J\sum_{\langle ij \rangle}Z_{i}Z_{j}-B\sum_{i}X_i,
\end{equation}
where the first sum runs over all nearest neighbor pairs. For the $1d$ case, the system is gapped as long as $J\neq B$ but gapless when $J=B$. In Ref.~\cite{Carleo602}, Carleo and Troyer demonstrated that the RBM state works very well in finding the ground state of the model. By minimizing the energy $E(\Omega)=\langle\Psi(\Omega)|H_{tIsing}|\Psi(\Omega)\rangle / \langle\Psi(\Omega)|\Psi(\Omega)\rangle$ with respect to the network parameters $\Omega$ using the improved gradient-descent optimization, they showed that the RBM states achieve an arbitrary accuracy for both $1d$ and $2d$ systems.

\paragraph{Antiferromagnetic Heisenberg model.}\textemdash The antiferromagnetic Heisenberg model is of the form
\begin{equation}\label{}
H=J\sum_{\langle ij\rangle}\mathbf{S}_i\mathbf{S}_j, (J>0)
\end{equation}
where the sum runs over all nearest neighbor pairs. In Ref.~\cite{Carleo602}, the calculation of the model is performed for the $1d$ and $2d$ systems using the RBM states. The accuracy of the neural network ansatz turns out to be much better than the traditional spin-Jastrow ansatz \cite{Jastrow1955} for the $1d$ system. The $2d$ system is harder, and more hidden neurons are needed to reach a high accuracy. In Ref.~\cite{Nomura2017}, a combined approach is presented; the RBM architecture was combined with a conventional variational Monte Carlo method with paired-product (geminal) wave functions to calculate the ground-state energy and ground state. They showed that the combined method has a higher accuracy than that achieved by each method separately.

\paragraph{$J_1$-$J_2$ Heisenberg model.}\textemdash The $J_1$-$J_2$ Heisenberg model (also known as the frustrated Heisenberg model) is of the form
\begin{equation}\label{}
H=J_1\sum_{\langle ij\rangle}\mathbf{S}_i\mathbf{S}_j+J_2\sum_{\langle\langle ij \rangle \rangle}\mathbf{S}_i\mathbf{S}_j,
\end{equation}
where the first sum runs over all nearest neighbor pairs and the second sum runs over the next-nearest-neighbor pairs. Cai and Liu \cite{Cai2018} produced expressions of the neural network states in this model using the feed-forward neural networks. They used the variational Monte Carlo method to find the ground state for the $1d$ system and obtained precisions to $\sim O(10^{-3})$. Liang and colleagues \cite{Liang2018} investigated the model using the convolutional neural network, and showed that the precision of the calculation based on convolutional neural network exceeds the string bond state calculation.

\paragraph{Hubbard model.}\textemdash The Hubbard model is a model of interacting particles on a lattice and endeavors to capture the phase transition between conductors and insulators. It has been used to describe superconductivity and cold atom systems. The Hamiltonian is of the form
\begin{equation}\label{}
H=-t\sum_{\langle ij\rangle,\sigma}(\hat{c}^{\dagger}_{i,\sigma}\hat{c}_{j,\sigma}
+\hat{c}^{\dagger}_{j,\sigma}\hat{c}_{i,\sigma})
+U\sum_{i}\hat{n}_{i,\uparrow}\hat{n}_{i,\downarrow},
\end{equation}
where the first term accounts for the kinetic energy and the second term the potential energy; $c^{\dagger}_{i,\sigma}$ and $c_{i,\sigma}$ denote the usual creation and annihilation operators, with $\hat{n}_{i,\sigma}=c^{\dagger}_{i,\sigma}c_{i,\sigma}$. The phase diagrams of the Hubbard model have not been completely determined yet. In Ref.~\cite{Nomura2017}, Nomura and colleagues numerically analyzed the ground state energy of the model by combining the RBM and the pair product states approach. They showed numerically that the accuracy of the calculation surpasses the many-variable variational Monte Carlo approach when $U/t=4,8$. A modified form of the model, described by the Bose\textendash Hubbard Hamiltonian, was studied in Ref.~\cite{saito2017solving} using a feed-forward neural network. The result is in good agreement with the calculation given by an exact diagonalization and the Gutzwiller approximation.

Here we briefly mention several important examples of numerical calculations of many-body physical systems. Numerous other numerical works concerning many different physical models have appeared. We refer the interested readers to e.g., Refs.~\cite{monterola2001solving,monterola2003solving,caetano2011using,manzhos2009improved,Carleo602,Rebentrost2014,Dunjko2016,Monras2017,carrasquilla2017machine,Deng2017,Deng2017a,jia2018efficient,gao2017efficient,Cai2018,Liang2018,Nomura2017,saito2017solving}

\section{Density operators represented by neural network}
\label{sec:density}
\subsection{Neural network density operator}
In realistic applications of quantum technologies, the states that we are concerned about are often mixed because the system is barely isolated from its environment. The mixed states are mathematically characterized by the density operator $\rho$ which is (i) Hermitian $\rho^{\dagger}=\rho$; (ii) positive semi-definite $\langle \Psi |\rho|\Psi\rangle\geq0$ for all $|\Psi \rangle$; and (iii) trace one $\mathrm{Tr}\rho=1$. The pure state $|\Psi\rangle$ provides a representation of the density operator $\rho_{\Psi}=|\Psi\rangle\langle\Psi|$ and the general mixed states are non-coherent superpositions (classical mixture) of pure density operators. Let us consider the situation for which the physical space of the system is $\mathcal{H}_S$ with basis $v_1,\cdots, v_n$ and the environment space is $\mathcal{H}_E$ with basis $e_1,\cdots,e_m$. For a given mixed state $\rho_S$ of the system, then if we take into account the effect of the environment there is a pure state $|\Psi_{SE}\rangle=\sum_{\mathbf{v}}\sum_{\mathbf{e}} \Psi(\mathbf{v},\mathbf{e}) |\mathbf{v}\rangle|\mathbf{e}\rangle$ for which $\rho_S=\mathrm{Tr}_{E}|\Psi_{SE}\rangle \langle\Psi_{SE} |$. Every mixed state can be purified in this way.

In Ref.~\cite{Torlai2018}, Torlai and Melko explored the possibility of representing mixed states $\rho_S$ using the RBM. The idea is the same as that for pure states. We build a neural network with parameters $\Omega$, and for the fixed basis $|\mathbf{v}\rangle$, the density operator is given by the matrix entries $\rho(\Omega,\mathbf{v},\mathbf{v}')$, which is determined by the neural network. Therefore, we only need to map a given neural network with parameters $\Omega$ to a density operator as
\begin{equation}\label{}
\rho(\Omega)=\sum_{\mathbf{v},\mathbf{v}'}|\mathbf{v}\rangle \rho(\Omega,\mathbf{v},\mathbf{v}')\langle \mathbf{v}'|.
\end{equation}

To this end, the purification method of the density operators is used. The environment is now represented by some extra hidden neurons $e_1,\cdots,e_m$ besides the hidden neurons $h_1,\cdots,h_l$. The purification $|\Psi_{SE}\rangle$ of $\rho_S$ is now captured by the parameters of the network, which we still denote as $\Omega$, i.e.,
\begin{equation}\label{}
|\Psi_{SE}\rangle=\sum_{\mathbf{v}}\sum_{\mathbf{e}}
\Psi_{SE}(\Omega,\mathbf{v},\mathbf{e})|\mathbf{v}\rangle|\mathbf{e}\rangle.
\end{equation}
By tracing out the environment, the density operator also is determined by the network parameters
\begin{equation}\label{eq:SE}
\rho_{S}=\sum_{\mathbf{v},\mathbf{v}'}[\sum_{\mathbf{e}}\Psi_{SE}(\Omega,\mathbf{v},\mathbf{e})\Psi_{SE}^*(\Omega,\mathbf{v}',\mathbf{e})]|\mathbf{v}\rangle\langle\mathbf{v}'|.
\end{equation}

\begin{figure}
  \centering
  \includegraphics[width=6cm]{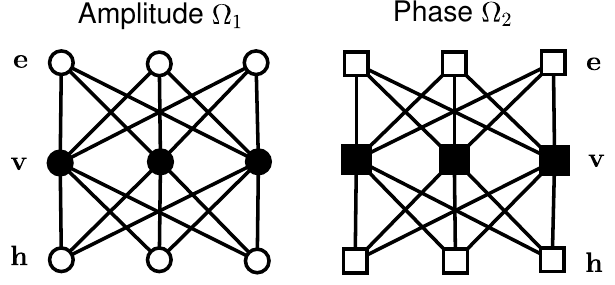}\\
  \caption{RBM construction of the latent space purification for a density operator.}\label{fig:density}
\end{figure}

To represent the density operators, Ref.~\cite{Torlai2018} takes the approach to represent the amplitude and phase of the purified state $|\Psi_{SE}\rangle$ by two separate neural networks. First, the environment units are embedded into the hidden neuron space, i.e., they introduced some new hidden neurons $e_1,\cdots,e_m$, which are fully connected to all visible neurons (See Figure~\ref{fig:density}). The parameters corresponding to the amplitude and phase of the wave function are now encoded in the RBM with two different sets of parameters. That is, the state $\Psi_{SE}(\Omega,\mathbf{v},\mathbf{e})=R(\Omega_1,\mathbf{a},\mathbf{v})e^{i\theta(\Omega_2,\mathbf{a},\mathbf{v})}$ with $\Omega=\Omega_1\cup\Omega_2$. $R(\Omega_1,\mathbf{a},\mathbf{v})$ and $\theta(\Omega_2,\mathbf{a},\mathbf{v})$ are both characterized by the corresponding RBM (this structure is called the latent space purification by authors).
In this way, the coefficients of the purified state $|\Psi_{SE}\rangle$ encoded by the RBM are
\begin{equation}\label{eq:density}
\Psi_{SE}(\Omega,\mathbf{v},\mathbf{e})=\sqrt{\frac{\sum_{\mathbf{h}}e^{-E(\Omega_1,\mathbf{v},\mathbf{h},\mathbf{e})}}{Z(\Omega_1)}}e^{i
\frac{\log \sum_{\mathbf{h}}e^{-E(\Omega_2,\mathbf{v},\mathbf{h},\mathbf{e})}}{2}},
\end{equation}
where $Z(\Omega_i)=\sum_{\mathbf{h}}\sum_{\mathbf{e}}\sum_{\mathbf{v}} e^{-E(\Omega_i,\mathbf{v},\mathbf{h},\mathbf{e})}$ is the partition function corresponding to $\Omega_i$. The density operator can now be obtained from Equation~(\ref{eq:SE}).

\subsection{Neural network quantum state tomography}
Quantum state tomography aims to identify or reconstruct an unknown quantum state from a dataset of experimental measurements. The traditional exact brute-force approach to quantum state tomography is only feasible for systems with a small number of degress of freedom otherwise the demand on computational resources is high. For pure states, the compressed sensing approach circumvents the experimental difficulty and requires only a reasonable number of measurements \cite{Gross2010}. The MPS tomography works well for states with low entanglement \cite{cramer2010efficient,lanyon2017efficient}. For general mixed states, the efficiency of the permutationally invariant tomography scheme based on the internal symmetry of the quantum states is low \cite{toth2010}. Despite all the progress, the general case for quantum state tomography is still very challenging.

The neural network representation of quantum states provides another approach to state tomography. Here we review its basic idea. For clarity (although there will be some overlap), we discuss its application to pure states and mixed states separately.

From the work by Torlai and colleagues, \cite{torlai2017many} for a pure quantum state, the neural network tomography works as follows. To reconstruct an unknown state $|\Psi\rangle$, we first perform a collection of measurements $\{\mathbf{v}^{(i)}\}$, $i=1,\cdots,N$ and therefore obtain the probabilities $p_{i}(\mathbf{v}^{(i)})=|\langle\mathbf{v}^{(i)}|\Psi\rangle|^2$. The aim of the neural network tomography is to find a set of RBM parameters $\Omega$ such that the RBM state $\Phi(\Omega,\mathbf{v}^{(i)})$ mimics the probabilities $p_{i}(\mathbf{v}^{(i)})$ as closely as possible in each basis. This can be done in neural network training by minimizing the distance function (total divergence) between $|\Phi(\Omega,\mathbf{v}^{(i)})|^2$ and $p_{i}(\mathbf{v}^{(i)})$. The total divergence is chosen as
\begin{equation}\label{}
D(\Omega)=\sum_{i=1}^N D_{KL}[|\Phi(\Omega,\mathbf{v}^{(i)})|^2|p_i(\mathbf{v}^{(i)})],
\end{equation}
where $D_{KL}[|\Phi(\Omega,\mathbf{v}^{(i)})|^2|p_i(\mathbf{v}^{(i)})]$ is the Kullback\textendash Leibler (KL) divergence in basis $\{\mathbf{v}^{(i)}\}$.

Note that to estimate the phase of $|\Psi\rangle$ in the reference basis, a sufficiently large number of measurement bases should be included. Once the training is completed, we get the target state $|\Phi(\Omega)\rangle$ in the RBM form, which is the reconstructed state for $|\Psi\rangle$. In Ref.~\cite{torlai2017many}, Torlai and colleagues test the scheme for the W-state, modified W state with local phases, Greenberger\textendash Horne\textendash Zeilinger and Dicke states, and also the ground states for the transverse-field Ising model and XXZ model. They find the scheme is very efficient and the number of measurement bases usually scales only polynomially with system size.

The mixed state case is studied in Ref.~\cite{Torlai2018} and is based on the RBM representations of the density operators. The core idea is the same as for the pure state; that is, to reconstruct an unknown density operator $\rho$, we need to build an RBM neural network density $\sigma(\Omega)$ with RBM parameter set $\Omega$. Before training the RBM, we must perform a collection of measurements $\{\mathbf{v}^{(i)}\}$ and obtain the corresponding probability distribution $p_i(\mathbf{v}^{(i)})=\langle\mathbf{v}^{(i)}|\rho|\mathbf{v}^{(i)}\rangle$. The training process involves minimizing the total divergence between the experimental probability distribution and the probability distribution calculated from the test RBM state $\sigma(\Omega)$. After the training process, we obtain a compact RBM representation of the density operator $\rho$, which may be used to calculate the expectation of the physical observable. Neural network state tomography is efficient and accurate in many cases. It provides a good supplement to the traditional tomography schemes.

\section{Entanglement properties of neural network states}
\label{sec:entanglement}
The notion of entanglement is ubiquitous in physics. To understand the entanglement properties of the many-body state is a central theme in both condensed matter physics and quantum information theory. Tensor network representations of quantum states have an important advantage in that entanglement can be read out more easily. Here, we discuss the entanglement properties of the neural network states for a comparison with tensor networks.

For a given $N$-particle quantum system in state $|\Psi\rangle$, we divide the $N$ particles into two groups $\mathcal{A}$ and $\mathcal{A}^c$. With this bipartition, we calculate the R\'{e}nyi entanglement entropy $S_{R}^{\alpha}(\mathcal{A}):=\frac{1}{1-\alpha}\log \mathrm{Tr}\rho_{\mathcal{A}} ^{\alpha}$, which characterizes the entanglement between $\mathcal{A}$ and $\mathcal{A}^c$, where $\rho_{\mathcal{A}}=\mathrm{Tr}_{\mathcal{A}^c}(|\Psi\rangle\langle\Psi|)$ is the reduced density matrix. If the R\'{e}nyi entanglement entropy is nonzero, then $\mathcal{A}$ and $\mathcal{A}^c$ are entangled.
\begin{figure}
  \centering
  \includegraphics[width=6cm]{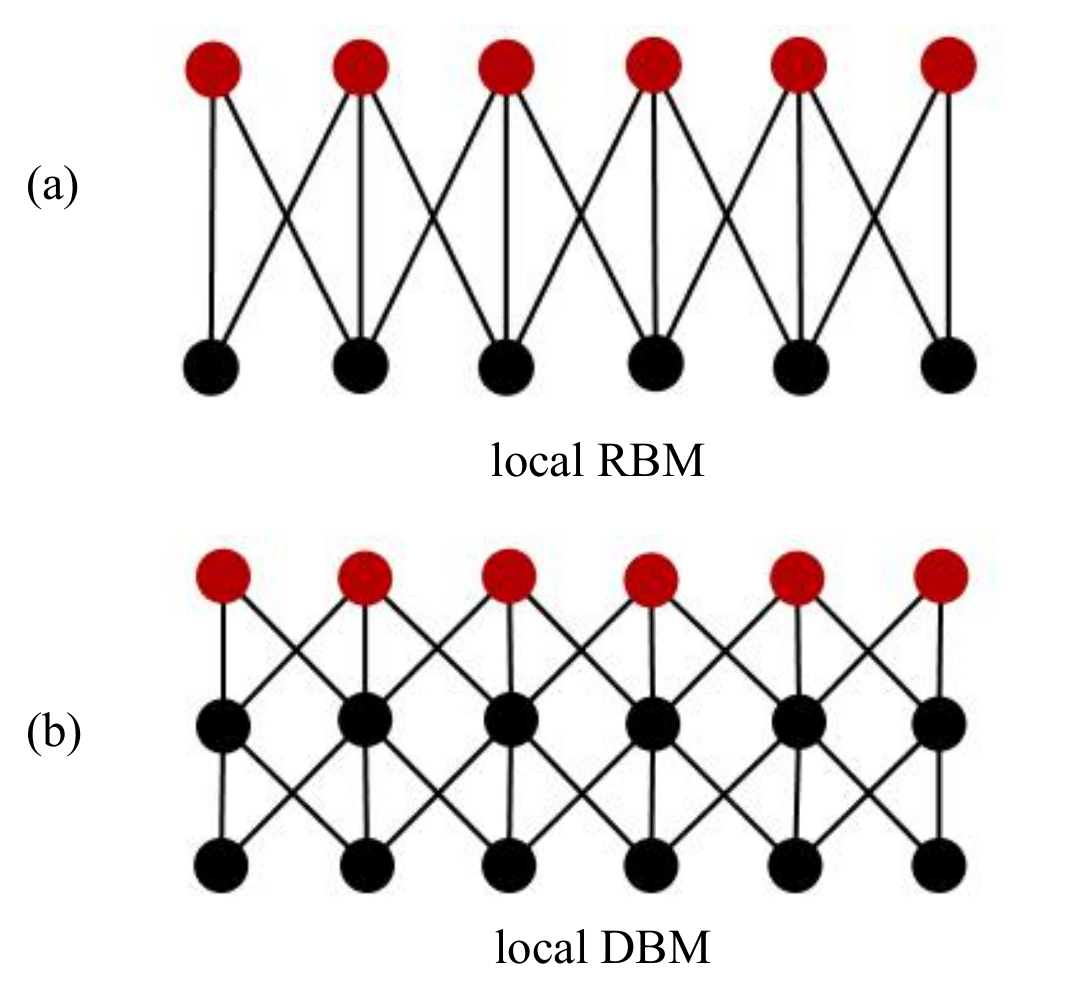}\\
  \caption{Example of (a) a local RBM state and (b) a local DBM state.}\label{fig:local}
\end{figure}

The entanglement property is encoded in the geometry of the contraction patterns of the local tensors for tensor network states. For neural network states, it was shown that the entanglement is encoded in the connecting patterns of the neural networks \cite{Deng2017,Huang2017,Chen2018,You2018,jia2018DBM}. For RBM states, Deng, Li, and Das Sarma \cite{Deng2017} showed that locally connected RBM states obey the entanglement area law, see Figure~\ref{fig:local}(a) for an illustration of a local RBM state. Nonlocal connections result in the volume-law entanglement of the states \cite{Deng2017}. We extended this result for any BM, showing that by cutting the intra-layer connection and adding hidden neurons, any BM state may be reduced to a DBM state with several hidden layers. Then using the folding trick, folding the odd layers and even layers separately, every BM is reduced into a DBM with only two hidden layers \cite{gao2017efficient,jia2018DBM}. Then we showed that the locally connected DBM states obey the entanglement area law, and the DBM with nonlocal connections possess volume-law entanglement \cite{jia2018DBM}, see Figure~\ref{fig:local}(b) for an illustration of a local DBM state.

The relationship between the BM and tensor network states was investigated in Refs.~\cite{gao2017efficient,Huang2017,Chen2018}, and some algorithmic way of transforming an RBM state into a MPS was given in Ref.~\cite{Chen2018}. The capability to represent tensor network states using the BM was investigated in \cite{gao2017efficient,Huang2017} from a complexity theory perspective.

One final aspect is realizing the holographic geometry-entanglement correspondence using BM states \cite{You2018,jia2018DBM}. When proving the entanglement area law and volume law of the BM states, the concept of locality must be introduced, this means that we must introduce a geometry between neurons. This geometry results in the entanglement features of the state. When we try to understand the holographic entanglement entropy, we first tile the neurons in a given geometry and then make it learn from data. After the learning process is done, we can see the connecting pattern of the neural network and analyze the corresponding entanglement properties, which have a direct relationship to the given geometry, such as the signs of the space curvature.

Although much progress on the entanglement properties of neural network states has been made, we still know very little about it. The entanglement features of neural networks other than the BM have not been investigated at all and remain to be explored in future work.

\section{Quantum computing and neural network states}
\label{sec:QC}
There is another crucial application of neural network states, namely, classical simulation of quantum computing, which we briefly review in this section. It is well-known that quantum algorithms can provide exponential speedup over some of the best known classical algorithms for many problems such as factoring integers \cite{Nielsen2010}. Quantum computers are being actively developed of late, but one crucial problem, known as quantum supremacy \cite{preskill2012quantum}, emerges naturally. Quantum supremacy concerns the potential capabilities of quantum computers that classical computers practically do not have and the resources required to simulate quantum algorithms using a classical computer. Studies of classical simulations of quantum algorithms can also guide us to understand what are the practical applications of the quantum computing platforms developed recently in different laboratories. Here we introduce the approach to simulating quantum circuits based on the neural network representation of quantum states.

Following Ref.~\cite{gao2017efficient}, we first discuss how to simulate quantum computing via DBM states, since in the DBM formalism, all operations can be written out analytically. A general quantum computing process can be loosely divided into three steps: (i) initial state preparation, (ii) applying quantum gates, and (iii) measuring the output state. For the DBM state simulation in quantum computing, the initial state is first represented by a DBM network. We are mainly concerned in how to apply a universal set of quantum gates in the DBM representations. As we shall see, this can be achieved by adding hidden neurons and weighted connections. Here the universal quantum gates is chosen as single-qubit rotation around $\hat{z}$-axis $Z(\theta)$, the Hadamard gate $H$, and controlled rotations around the $\hat{z}$-axis $CZ(\theta)$ \cite{Barenco1995}.

We continue still to denote the calculating basis by $|\mathbf{v}\rangle$; the input state is then represented by DBM neural network as $\Psi_{\mathrm{in}}(\mathbf{v},\Omega)=\langle \mathbf{v}|\Psi_{\mathrm{in}}(\Omega)\rangle$. To simulate the circuit quantum computing, characterized by unitary transform $U_{C}$, we need to devise strategies so that we can apply all the universal quantum gates to achieve the transform,
\begin{equation}\label{eq:QC}
\langle \mathbf{v}|\Psi_{\mathrm{in}}(\Omega)\rangle\overset{DBM}{\to}
\langle \mathbf{v}|\Psi_{\mathrm{out}}(\Omega)\rangle
=\langle \mathbf{v}|U_C|\Psi_{\mathrm{in}}(\Omega)\rangle.
\end{equation}

Let us first consider how to construct the Hadamard gate operation
\begin{equation}\label{}
  H|0\rangle=\frac{1}{\sqrt{2}}(|0\rangle+|1\rangle),
  H|0\rangle=\frac{1}{\sqrt{2}}(|0\rangle-|1\rangle).
\end{equation}
If $H$ acts on the $i$-th qubit of the system, we can then represent the operation in terms of the coefficients of the state,
\begin{align}\label{}
\Psi(\cdots v_i\cdots)&\overset{H}{\to}\Psi'(\cdots v'_i\cdots)\nonumber\\
=&\sum_{v_i=0,1}\frac{1}{\sqrt{2}}(-1)^{v_iv'_i}\Psi(\cdots v_i\cdots).
\end{align}
In DBM settings, it is clear now that the Hadamard DBM transform of the $i$-th qubit adds a new visible neuron $v'_i$, which replaces $v_i$, and another hidden neuron $H_i$ and $v_i$ now becomes a hidden neuron. The connection weight is given by $W_{H}(v,H_i)=\frac{i\pi}{8}-\frac{\ln}{2}-\frac{i\pi v}{2}-\frac{i\pi H_i}{4}+i\pi v H_i$, where $v=v_i,v'_i$. We easily check that $\sum_{H_i=0,1} e^{W_{H}(v_i,H_i)+W_{H}(v'_i,H_i)}=\frac{1}{\sqrt{2}}(-1)^{v_iv'_i}$, which completes the construction of the Hadamard gate operation.

The $Z(\theta)$ gate operation,
\begin{equation}\label{}
Z(\theta)|0\rangle=e^{\frac{-i\theta}{2}}|0\rangle,\qquad
Z(\theta)|1\rangle=e^{\frac{i\theta}{2}}|1\rangle,
\end{equation}
can be constructed similarly. We can also add a new visible neuron $v'_i$ and a hidden neuron $Z_{i}$, and $v_i$ becomes a hidden neuron that should be traced. The connection weight is given be $W_{Z(\theta)}(v,Z_i)=-\frac{\ln 2}{2}+\frac{i\theta v}{2}+i\pi v Z_i$ where $v=v_i,v'_i$. The DBM transform of the controlled $Z(\theta)$ gates is slightly different from single qubit gates because it is a two-qubit operation acting on $v_i$ and $v_j$. To simplify the calculation, we give here the explicit construction for $CZ$. This can be done by introducing a new hidden neuron $H_{ij}$, which connects both $v_i$ and $v_j$ with the same weights as those given by the Hadamard gate. In summary, we have
\begin{equation}\label{}
\Qcircuit @C=1em @R=.7em {& \gate{H} & \qw}
\Leftrightarrow\f{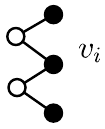}\overset{DBM}{\to}\f{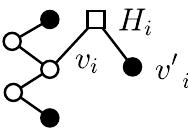}\, ,
\end{equation}

\begin{equation}\label{}
\Qcircuit @C=1em @R=.7em {& \gate{Z(\theta)} & \qw}
\Leftrightarrow\f{H1.pdf}\overset{DBM}{\to}\f{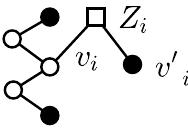}\, ,
\end{equation}

\begin{align}\label{}
\begin{array}{cccccc}
\Qcircuit @C=1em @R=.7em {& \ctrl{1} & \qw \\& \gate{Z} & \qw}
\end{array}
\Leftrightarrow\f{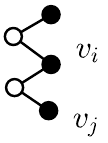}\overset{DBM}{\to}\f{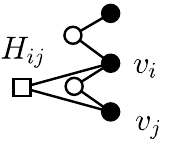}.
\end{align}

Note that we can also achieve the quantum gate $H$ in the DBM setting by adding directly a new visible neuron $v'_i$ and connecting it with the (hidden) neuron $v_i$. $Z(\theta)$ can also be realized in the DBM setting by changing the bias of the visible neuron $v_i$. We choose the method presented above simply to make the construction clearer and more systematic.

The above protocol based on the DBM is an exact simulation but has a drawback in that the sampling of the DBM quickly becomes intractable with increasing depth of the circuit because the gates are realized by adding deep hidden neurons. In contrast, RBMs are easier to train; a simulation based on the RBM has already been developed \cite{jonsson2018neural}. The basic idea is the same as the DBM approach, the main difference being that Hadamard gate cannot be exactly simulated in the RBM setting. In Ref.~\cite{jonsson2018neural}, the authors developed the approximation method to simulate the Hadamard gate operation. The RBM realizations of $Z(\theta)$ and $CZ(\theta)$ are achieved by adjusting the bias and introducing a new hidden neuron and weighted connections, respectively.

\section{Concluding remarks}
\label{sec:conclusion}

In this work, we discussed aspects of the quantum neural network states. Two important kinds of neural networks, feed-forward and stochastic recurrent, were chosen as examples to illustrate how neural networks can be used as a variational ansatz state of quantum many-body systems. We reviewed the research progress on neural network states. The representational power of these states was discussed and entanglement features of the RBM and DBM states reviewed. Some applications of quantum neural network states, such as quantum state tomography and classical simulations of quantum computing, were also discussed.

In addition to the foregoing, we present some remarks on the main open problems regarding quantum neural network states.
\begin{itemize}
  \item One crucial problem is to explain why the neural network works so well for some special tasks. There should be deep reasons for this. Understanding the mathematics and physics behind the neural networks may help to build many other important classes of quantum neural network states and guide us in applying the neural network states to different scientific problems.
  \item Although the BM states have been studied from various aspects, many other neural networks are less explored in regard to representing quantum states both numerically and theoretically. This raises the question whether other networks can also efficiently represent quantum states, and what are the differences between these different representations?
  \item Developing the representation theorem for the complex function is also a very important topic in quantum neural network states. Because we must build the quantum neural network states from complex neural networks, as we have discussed, so it is important to understand the expressive power of the complex neural network.
  \item Having a good understanding of entanglement features is of great importance in understanding quantum phases and the quantum advantage over some information tasks. Therefore, we can also ask if there is an easy way to read out entanglement properties from specific neural networks such as the tensor network.
\end{itemize}

We hope that our review of the quantum neural network states inspires more work and exploration of the crucial topics highlighted above.

\begin{acknowledgments}
Z.-A. Jia thanks Zhenghan Wang and hospitality of Department of Mathematics of UCSB. He also acknowledges Liang Kong and Tian Lan for discussions during his stay in Yau Mathematical Science Center of Tsinghua University, and he also benefits from the discussion with Giuseppe Carleo during the first international conference on ``Machine Learning and Physics" at IAS, Tsinghua University. This work was supported by the Anhui Initiative in Quantum Information Technologies (Grant No. AHY080000).
\end{acknowledgments}

\bibliographystyle{apsrev4-1-title}
\end{document}